# All-optical control of long-lived nuclear spins in rare-earth doped nanoparticles

D. Serrano[1], J. Karlsson[1], A. Fossati[1], A. Ferrier[1,2] and P. Goldner[1]

[1] PSL Research University, Chimie ParisTech, CNRS, Institut de Recherche de Chimie Paris

75005, Paris, France

[2] Sorbonne Universités, UPMC Université Paris 06, 75005, Paris, France

**Nanoscale systems offer key capabilities for quantum technologies that include single qubit control and readout[1], multiple qubit gate operation[2,3], extremely sensitive and localized sensing and imaging[4], as well as the ability to build hybrid quantum systems[5]. To fully exploit these functionalities, multiple degrees of freedom are highly desirable[6]: in this respect, nanoscale systems that coherently couple to light and possess spins, allow for storage of photonic qubits or light-matter entanglement together with processing capabilities. In addition, all-optical control of spins can be possible for faster gate operations[7] and higher spatial selectivity compared to direct RF excitation. Such systems are therefore of high interest for quantum communications and processing. However, an outstanding challenge is to preserve properties, and especially optical and spin coherence lifetimes, at the nanoscale[8,9]. Indeed, interactions with surfaces related perturbations strongly increase as sizes decrease, although the smallest objects present the highest flexibility for integration with other systems.**

Here, we demonstrate optically controlled nuclear spins with long coherence lifetimes ($T_2$) in rare earth doped nanoparticles. We observed spins echoes and measured $T_2$ of 2.9 ± 0.3 ms at 5 K and under a magnetic field of 9 mT, a value comparable to those obtained in bulk single crystals[10]. Moreover, we achieve, for the first time, spin $T_2$ extension using all-optical spin dynamical decoupling and observe high fidelity between excitation and echo phases.

Rare-earth doped nanoparticles are thus the only reported nano-materials in which optically controlled spins with millisecond coherence lifetimes have been observed. These results open the way to providing quantum light-atom-spin interfaces with long storage time within hybrid architectures.

**Keywords:** rare-earths, nanomaterials, quantum technologies, spin, dynamical decoupling.

Quantum systems with spin qubits that can be optically controlled allow efficient qubit initialization and readout, and quantum gate operations[3]. Moreover, photonic quantum states can be mapped to and/or entangled with spin qubits for storage and processing[11-13]. Such schemes are investigated in solid-state systems like colour centres in diamond, quantum dots in semi-conductors, and rare earth doped crystals. Targeted applications include quantum memories for light[11,14,15] or microwave photons[16], quantum processors[3] and sensors[4]. Crucial advances are expected at the nanoscale: light-matter interactions can be strongly enhanced using nano- or micro-cavities[17,18], leading to more efficient detection and control at the single centre level. Different quantum systems could be coupled together to build hybrid systems with an optical interface[5]. Optical control of spins can also be useful in nanoscale systems. It can be faster than direct RF excitation because it takes advantage of strong optical transitions[7], while ensuring spatial selectivity because of light much shorter wavelength. It may also lead to simpler fabrication of devices by avoiding incorporating antennas in proximity to the spins.

However, coherence lifetimes are often significantly shortened in nano-materials[8,9], impairing their use for quantum technologies. Indeed, surface effects, and/or high concentration of defects or impurities due to the synthesis process can cause strong dephasing mechanisms[8]. The latter can be partially cancelled in nanostructures embedded in bulk crystals[17,19], although freestanding nanoparticles are better suited for building systems from independent parts. As an example, nanoparticles can be integrated in high-finesse optical micro-cavities[18] or coupled through an active substrate on which they are deposited[20]. Here, we investigate nuclear spin coherences in rare earth doped nanoparticles at low temperatures using all-optical techniques. These materials have unique properties for nanoscale systems, showing narrow optical linewidths, down to 45 kHz at 1.3 K, and limited spectral diffusion[21]. This is favourable to coupling to high finesse optical cavities and using electric dipole-dipole interactions for quantum gate implementation.

Experiments were carried out on 0.5 % $Eu^{3+}:Y_2O_3$ nanoparticles of 400 ± 80 nm composed of 130 ± 10 nm crystallites obtained by homogeneous precipitation and high temperature annealing[22]. They were placed in a cryostat in the form of a powder and excited by laser pulses as described in the Methods section. With a nuclear spin I=5/2, the $^{151}$Eu isotope presents three doubly degenerated ground-state nuclear spin levels at zero magnetic field (Fig. 1a). To probe the ±1/2↔±3/2 hyperfine transition, the thermally distributed ground-state population was first initialized by optical pumping to the ±1/2 level for a subset of ions within the inhomogeneously broadened optical absorption line (Fig. 1b). Spin coherent states were subsequently created and rephased following an all-optical spin echo sequence[23], using two-color pulses at frequencies $\omega_1$ and $\omega_2$ (Fig. 1c). A weak single-frequency pulse was applied at time $2\tau$ with frequency $\omega_2$ to convert the spin coherence into an optical coherence at $\omega_1$. This resulted in a beating at $\omega_2-\omega_1$ on the photodiode signal that was revealed with a signal to noise ratio of about 10 by a fast Fourier transform (FFT) as displayed in Fig. 1d.

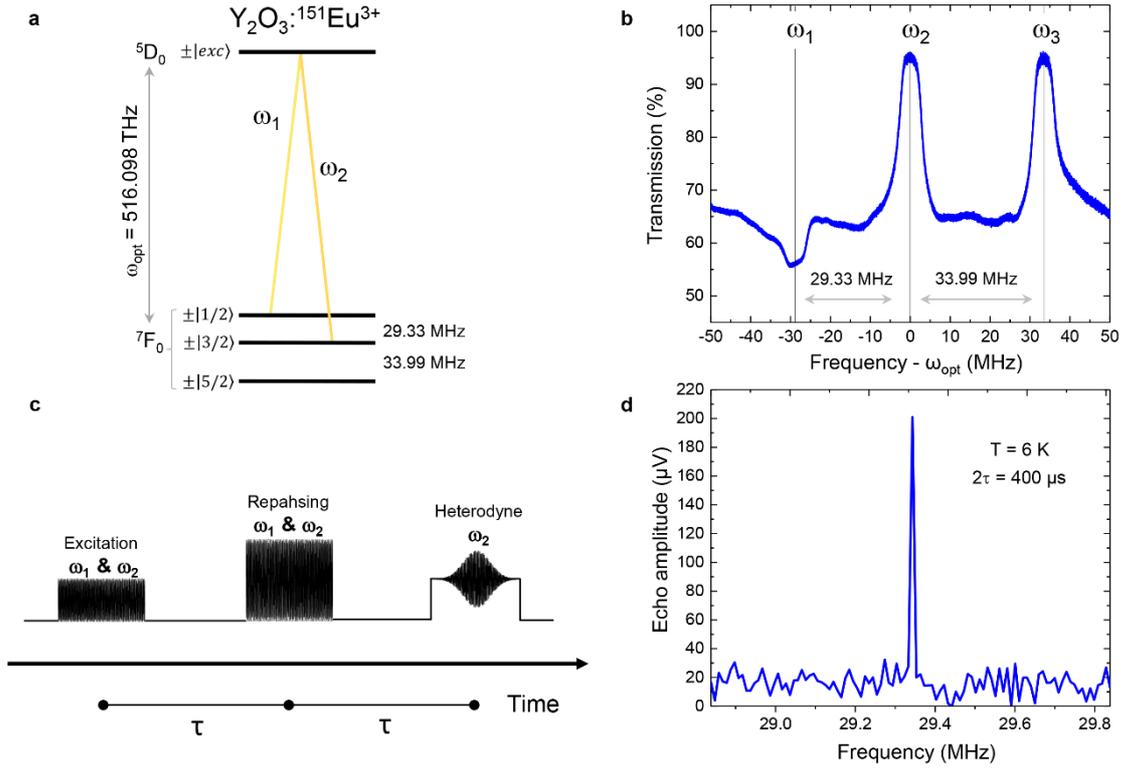

**Figure 1: All-optical nuclear spin coherence investigation in $^{151}$Eu$^{3+}$ doped Y$_2$O$_3$ nanoparticles. a.** $^{151}$Eu$^{3+}$ ground-state hyperfine structure in Y$_2$O$_3$. Two-color laser pulses (at $\omega_1$ and $\omega_2$ frequencies) resonant with the $^7F_0 \rightarrow {}^5D_0$ transition at 580.883 nm create coherent states between the ±1/2 and ±3/2 nuclear spin levels. |±exc⟩ represents the excited state hyperfine levels. **b.** Optical transmission spectrum after optical pumping. Ground-state population initialization to ±1/2 is corresponds to a lower transmission at $\omega_1$. High transmission (~95%) at 0 ($\omega_2$) and 33.99 MHz ($\omega_3$) evidence efficient population depletion in the ±3/2 and ±5/2 levels. $\omega_{opt}$ = 516.098 THz (580.883 nm). **c.** All-optical spin echo sequence with heterodyne detection. Each sequence is preceded by optical pumping and followed by chirped pulses to reset the spin population to equilibrium **d.** Fast Fourier transform of the heterodyne signal revealing the spin echo at 29.34 MHz.

This sequence was first used to determine the inhomogeneous broadening of the ±1/2↔±3/2 transition, 107±8 kHz (Fig. 2a). This value, identical to that reported on Y$_2$O$_3$:Eu$^{3+}$ bulk crystals[24] and ceramics[25], reflects the high crystalline quality of the particles. The decay of the spin echo amplitude as a function of the increasing pulse separation reveals a coherence lifetime of T$_2$ = 1.3 ± 0.2 ms (Fig. 2b), corresponding to a homogeneous linewidth $\Gamma_h = (\pi T_2)^{-1}$ of 250 Hz. Remarkably, this spin coherence lifetime is only one order of magnitude lower compared to Eu$^{3+}$:Y$_2$O$_3$ bulk transparent ceramics (T$_2$ = 12 ms[25]) and Eu$^{3+}$:Y$_2$SiO$_5$ bulk single crystals (T$_2$ = 19 ms[10]). In contrast, the nanoparticles' optical coherence lifetime is two orders of magnitude lower than the bulk values (T$_{2opt}$ = 7 $\mu$s[21] versus T$_{2opt}$ = 1.1 ms[26]). Thus, the spin coherence is much more preserved when scaling down in size than the optical coherence. This is consistent with a previous analysis in which we proposed that optical dephasing is mainly due to perturbations related to surface electric charges[21]. These charges have however little influence on nuclear spin transitions as the ratio between optical and nuclear Stark coefficients is expected to be about 5 orders of magnitude[27]. This suggests that magnetic perturbations are responsible for the dephasing of the spin transition. Indeed, under a weak magnetic field, the homogeneous linewidth strongly decreases and reaches 110 Hz at 9 mT (Fig. 2c). This variation can be modelled by magnetic dipole-dipole (dd) interactions between $Eu^{3+}$ spins and defects carrying electron spins (Fig. 2c and Supporting Information (SI)). A small magnetic field reduces the dd interaction Hamiltonian to secular terms, which in turn reduces $Eu^{3+}$ spin frequency shifts due to defect spin flips and therefore dephasing[28]. Quantitative analysis was performed assuming that $Eu^{3+}$ spin dephasing is due to frequency shifts following a Gaussian distribution (SI). The inferred defect concentration, $6.4 \times 10^{17}$ cm$^{-3}$ or 25 ppm relative to Y, also indicates that spin $T_2$ could be increased in higher quality samples[28].

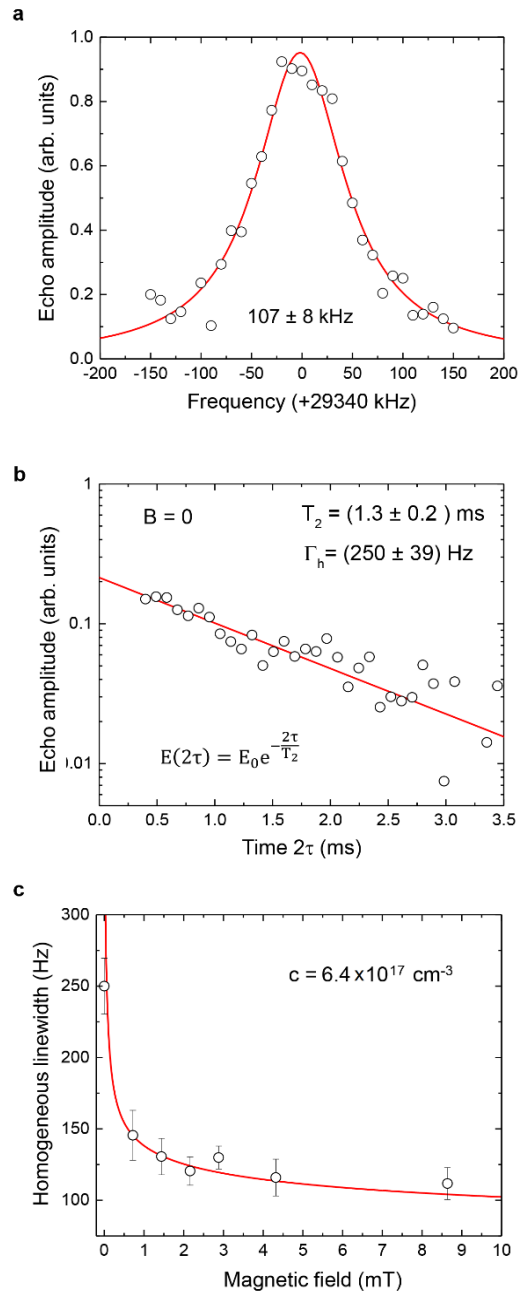

**Figure 2:** **$^{151}$Eu$^{3+}$ spin inhomogeneous and homogeneous linewidths a.** Inhomogeneous linewidth of the ±1/2↔±3/2 spin transition obtained by monitoring the spin-echo amplitude as a function of the frequency detuning $\omega_2 - \omega_1$ for a fixed time delay $2\tau$ of 400 µs (circles). Solid line: Lorentzian fit corresponding to a full width at half maximum of 107 ± 8 kHz. **b.** Spin echo decay at zero magnetic field. A single exponential fit yields a coherence lifetime $T_2$ of 1.3 ± 0.2 ms, corresponding to a $\Gamma_h$ = 250 Hz homogeneous linewidth. **c.** Homogeneous linewidth evolution under an applied external magnetic field. A drastic decrease in $\Gamma_h$ is observed for weak fields, corresponding to a coherence lifetime increasing from 1.3 ms to 2.9 ms (Fig. S5). Solid line: modelling by interactions with defects carrying electron spins at a concentration of 6.4 10$^{17}$ cm$^{-3}$.

A well-known approach to control dephasing is dynamical decoupling (DD)[29]. Here, a train of $\pi$ pulses is applied to refocus frequency shifts due to fluctuations that are slow compared to the pulse separation. This principle was applied but with $\pi$ pulses corresponding to two-color laser pulses instead of the usual radio-frequency (RF) ones[10]. To the best of our knowledge this the first demonstration of an all-optical spin dynamical decoupling. A crucial point for DD, is the phase coherence of the $\pi$ pulses. We achieved it by generating the two frequency shifted laser beams using a single acousto-optic modulator and having them spatially overlap (see Methods). This ensured a highly stable relative phase between the two lasers beams and therefore phase coherent excitation, rephasing, and detection of spins.

The CPMG (Carl-Purcell-Meiboom-Gill)[30] dynamical decoupling sequence used in our experiments is shown in Fig. 3a. Coherence lifetimes extended by DD, $T_{2DD}$, were determined by recording the spin echo amplitude vs. the total evolution time (n×$\tau_{DD}$). This is efficient in preserving coherences along the X axis of the Bloch sphere, but not those along the Y axis. This is due to the accumulation of errors in the $\pi$ pulse areas that have a larger effect for Y coherences than for X ones. In our powder, such pulse area errors are expected to be particularly high because of the random light scattering and orientation of the particles, which further increases the spread in spin Rabi frequencies. Indeed, significant increase in coherence lifetime over the two-pulse echo value of 1.3 ms were achieved only for X excitations (Fig. 3b). The $\pi$ pulse delay $\tau_{DD}$ was then varied, resulting in $T_{2DD}$ = 8.1 ± 0.6 ms for the optimal value $\tau_{DD}$ = 300 $\mu$s, a 6-fold increase compared to the two-pulse echo $T_2$ (Fig. 3c). $T_{2DD}$ variation with $\tau_{DD}$, shown in Fig. 3d, can be explained by a balance between short $\tau_{DD}$ delays implying a higher number of pulses during a given evolution time and therefore accumulating pulse areas errors, and long delays that are less efficient in refocusing fluctuations[10] (SI). We also noted that applying a field of 0.7 mT decreased $T_{2DD}$, in opposition to $T_2$ (Fig. 3e). This could be explained by an increase in pulse errors when the transition broadens under magnetic field.

We finally investigated the variation of spin echo phase as a function of the initial excitation phase in the 2 pulse and DD all-optical sequences (see Methods). They were found to be highly correlated, even for the DD case, in which a lower signal to noise ratio was achieved (Fig. 4 and S8). This confirmed the fully coherent character of the spins driving and detection. These experiments can also be considered as an optical to spin coherent storage, with the initial and final light fields at $\omega_1$ being input and output signals (Fig. 1c). The high correlations of Fig. 4 then indicate a high fidelity optical memory, an essential requirement towards an optical memory operating at the quantum level. In this respect, further investigations on the noise level introduced by all-optical DD will be important to assert the possibility of long time spin storage.

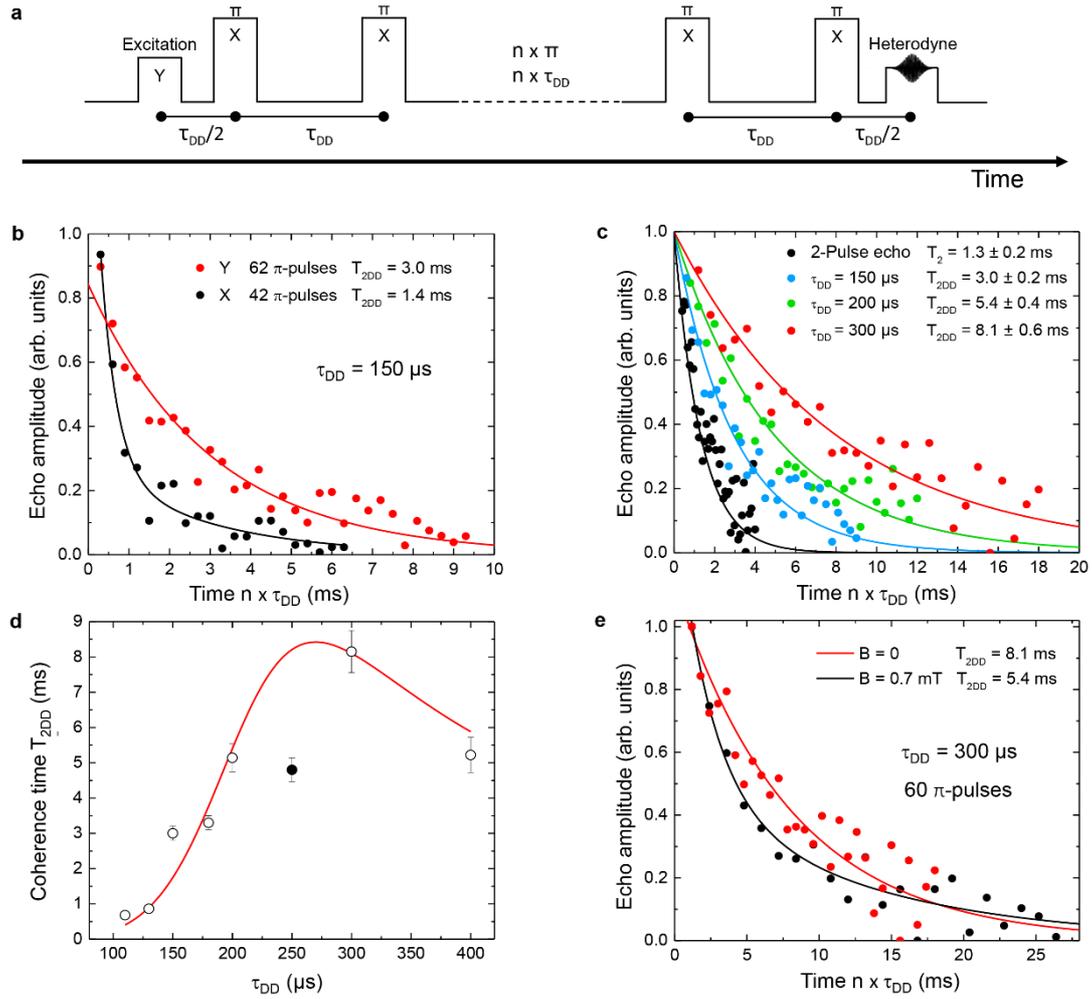

**Figure 3: All-optical dynamical decoupling a.** CPMG sequence with optical 2-color excitation and rephrasing π pulses, and heterodyne detection. The initial excitation pulse has a Y phase and the π pulses an X phase[22]. This is obtained by varying the relative phase between the two frequency components of the optical pulses. **b.** Echo decays (circles) for different initial phases. Lines: exponential fits. A much lower $T_{2DD}$ is observed for an X initial phase (~1.0 ms) than for an Y one (~3.0 ms). This is due to the accumulation of pulse errors and confirms that our DD sequence behaves as a CPMG one. **c.** Spin echo decays (circles) obtained for $\tau_{DD}$ = 150, 200 and 300 μs and n ≤ 60. Solid lines: exponential fits. **d.** Experimental (circles) and modelled (line) $T_{2DD}$ evolution as a function of $\tau_{DD}$. (see SI). The data point represented by the black circle was discarded for the fit **e.** Spin echo decays (circles) with and without a weak magnetic field. Solid line: exponential fits.

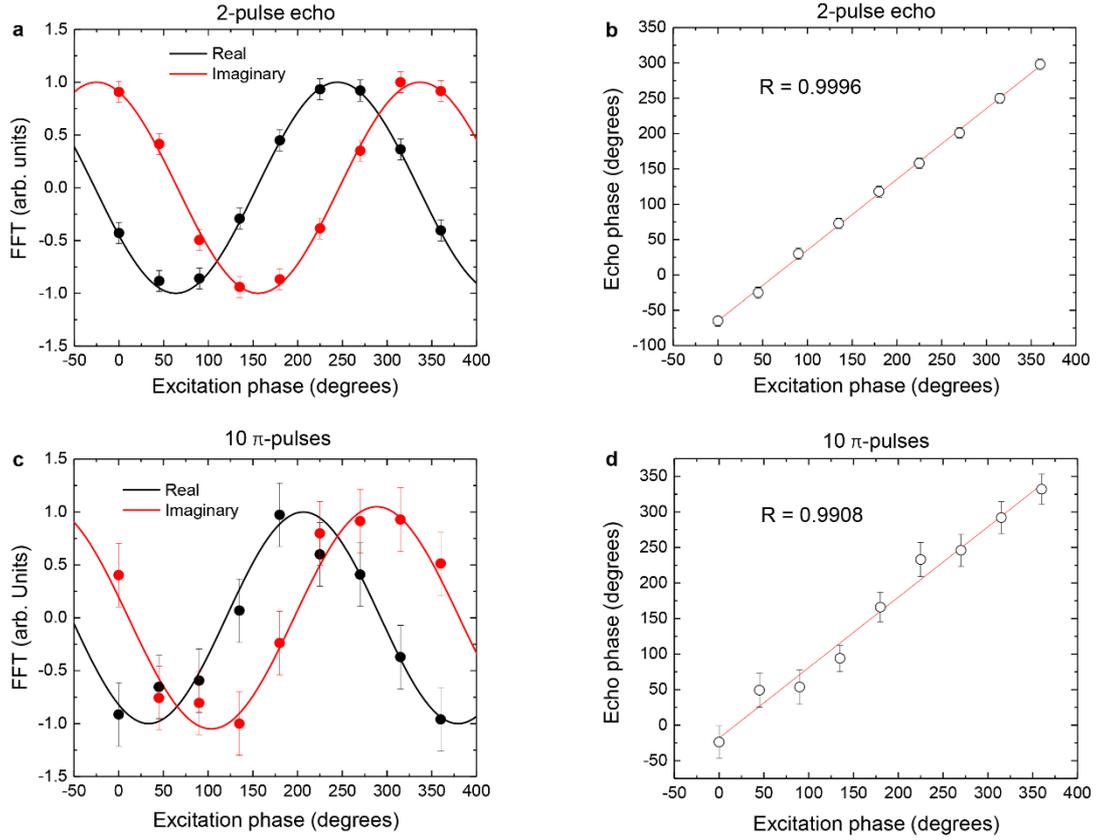

**Figure 4: Echo phase correlation a,c** Real and imaginary parts of the spin echo signal FFT (squares and circles) as a function of the excitation pulse phase for a two-pulse echo and a DD sequence ($\tau$ = 300 $\mu$s, $\tau_{DD}$ = 150 $\mu$s, n = 10). Solid lines: fit with a sine function. **b,d.** Echo pulse phase derived from FFT signals (a,c) as a function of the excitation pulse phase. Lines: linear fit, R correlation coefficient.

In conclusion, we have measured nuclear spin coherence lifetimes in the ms range in $Eu^{3+}$ doped $Y_2O_3$ nanoparticles using a fully-optical protocol. These values are remarkably close to bulk values and could be increased in samples with lower content of magnetic defects. Moreover, in single particles, $T_2$ could be increased by several orders of magnitude, by reducing pulses area errors in DD and using clock transitions that appear in europium and other rare earth ions under suitable magnetic fields[31]. These results open the way to nanoscale quantum light-matter-spin interfaces, useful for quantum memories with processing capabilities, hybrid opto-mechanical systems or coupling to optical micro-cavities. Nanoparticles doped with essentially any rare earth ion can also be synthesized in different size, shape and layered structures, as shown by their huge development as luminescent probes[32]. Although quantum grade materials are very demanding, our results suggest that rare earth ion doped nanoparticles could be an extremely versatile platform for nanoscale quantum technologies.

**Methods**

0.5% $Eu^{3+}$:$Y_2O_3$ nanoparticles with 400 ± 80 nm average diameter and 130 ± 10 nm crystallite size were synthesized by calcining monodispersed spherical particles of $Eu^{3+}$-doped yttrium basic carbonate ($Eu^{3+}$–Y(OH)$CO_3$·n$H_2O$) grown by homogeneous precipitation as described in reference[22]. Body-centered cubic Y2O3 structure (Ia-3 space group) was confirmed by X-ray diffraction and no evidence of other parasitic phases were found. The sample, consisting of ensembles of particles in form of loose powder was placed between two glass plates with a copper spacer (~500 µm thickness) inside a He bath cryostat (Janis SVT-200). The excitation was provided by a Sirah Matisse DS laser, with a linewidth of approximately 150 kHz and operating at 516.098 THz (580.883 nm vac.) The laser beam was first sent through a double pass acousto-optic modulator (AOM) with central frequency of 200 MHz (AA Optoelectronic MT200-B100A0, 5-VIS) followed by a single pass AOM (AA Optoelectronic MT110-B50A1-VIS) with a center frequency of 110 MHz. Both AOMs were driven by an arbitrary waveform generator (AWG) (Agilent N8242A) with 625 MS/s sampling rate. The two-color pulses, generated by the single pass AOM, were coupled to a single-mode fiber in order to ensure spatial overlapping. The overlapped beams were then focused onto the sample, within the cryostat, with a 75 mm focal length lens and the scattered light collected with a 5 mm focal length lens mounted directly behind the sample holder. An avalanche photo diode (APD) (Thorlabs 110 A/M) was used as detector. A scheme of the experimental setup is displayed in SI (Fig. S1). The sample temperature was monitored with a temperature sensor (Lakeshore DT-670) attached to the sample holder with thermally conducting grease and tuned by controlling the helium gas flow and the pressure inside the cryostat. The cryostat was operated in gaz mode to maintain a constant temperature of 5 K. Magnetic fields perpendicular the laser beam propagation axis were applied by means of Helmholtz coils sitting outside the cryostat.

*Two-pulse echo measurements*

Pulse areas in the two-pulse echo sequence were optimized to maximize the spin-echo signal. Data presented in this work were obtained with 100 µs long pulses and powers of the order of 120 mW. Although this input power is large compared to single crystal measurements, the scattering in the nanoparticles significantly reduces the input power incident in the sample. Lower excitation power was used for the heterodyne pulse (~14 mW). The inhomogeneous linewidth of the 29 MHz spin transition was measured by monitoring the spin-echo signal as a function of the frequency difference ($\omega_2$-$\omega_1$) in the two-color pulses for a fixed delay time τ. The transition linewidth was estimated by a Lorentzian lineshape fit within an incertitude interval which was derived from the experimental signal to noise ratio and the fit accuracy. The decay of the spin-echo signal with increasing τ was used to determine the nuclear spin coherence lifetime. The value was derived by single-exponential fit within an uncertainty also given by the experimental signal to noise ratio and the fit accuracy.

*Dynamical decoupling (DD) and phase correlations measurements*

DD experiments were carried out with 20-µs-long π pulses in order to access a large excitation bandwidth (about 50 kHz, half of the spin inhomogeneous linewidth) and short π-pulse separation times ($\tau_{DD}$). The preservation of the excitation phase coherence along the DD the sequence was confirmed by the observation of stable beating patterns from a photodiode at the output of the fiber for times exceeding 30 s. The spin-echo phase was directly derived from the real (*Re*) and imaginary (*Im*) parts of the spin-echo signal FFT as:

$$\theta_{echo} = \tan^{-1}\left(\frac{Im}{Re}\right) + n\pi$$

The error $\Delta\theta_{echo}$ was calculated by classical error propagation from the uncertainty associated to the real and imaginary FFT parts, $\Delta Re$ and $\Delta Im$. Those were estimated from the signal to noise ratio (SNR) in Fig. S6 and S7. As observed, the SNR is clearly larger in Fig. S7 due to the weaker spin-echo signal obtained after 10 π-pulses, corresponding to a total evolution time of 1.5 ms compared to the total evolution time of 600 µs in the two-pulse echo case.

**Acknowledgment**

This research work has received funding from the European Union's Horizon 2020 research and innovation programme under grant agreement No 712721 (NanQTech). We thank John Bartholomew and Hugues de Riedmatten for useful comments on the manuscript.

**Author's contribution**

JK developed the optical setup; DS, JK, AFo and PG performed the experiments; DS, JK, AFe and PG discussed, modelled and analyzed the results, DS and PG wrote the manuscript. All authors reviewed the manuscript.